# Time Extended Measurement of the Position of a Particle

Francesc S. Roig
Physics Department University of California Santa Barbara, CA 93106
(December 20, 2005)

The von Neumann interaction between a particle and an apparatus has been considered in the measurement of the position of a particle when the interaction lasts for a finite amount of time. When the measurement has finite duration, both the motion of the pointer and the particle influence the result of the measurement. Provided that the particle is in an eigenstate of its position at the start of the measurement, the pointer will indicate the arithmetic average between the initial and final position of the particle. Furthermore, the probability that the pointer will indicate a given average value is equal to the transition probability for the undisturbed particle to experience the change in position. If the initial state of the pointer is a narrow wavepacket, then for any initial state of the particle, the measurement yields, approximately, the undisturbed probability distribution for the position of the particle at the end of the measurement.

## 1. Introduction

Realistic measurements of the position of a particle do not take place instantaneously, rather the period of time of interaction between a measuring apparatus and the particle is extended compared to the particle's dynamical time, i.e., the time for significant changes in position under the given forces acting on the particle. For instance, the relaxation time of an electron in a copper wire is of the order of $10^{-15}$ s at room temperature. This sets the scale for dynamical time. By contrast, the time needed for an apparatus to measure the position of a particle, or any other dynamical quantity for that matter, is easily longer than this.

The usual von Neumann model [1] for the measurement of the position of a particle assumes that the interaction between the particle and the apparatus lasts for a very short time and is very strong. Thus the motion of the particle and the pointer can be neglected during the measurement process. If the position of the particle is $x$ and the momentum of the pointer is $P$ then the interaction Hamiltonian describing this interaction is given by

$$H_i = \frac{1}{T} f(t) x P \tag{1}$$

where $T$ is the very short time this interaction lasts and $f(t)$ is a dimensionless coupling function of time with compact support $[0,T]$. Just before the measurement, the state of the system is the product state $\psi_0(x, X) = \varphi_0(x)\Phi_0(X)$, where $\varphi_0(x)$ is the state of the particle and $\Phi_0(X)$ is a normalized wavepacked centered at $X = 0$ describing the state of the pointer, and $X$ is the pointer coordinate. If we assume that $T$ is very short, compared to any dynamical time scale for particle and pointer, then the interaction (1) by itself makes the system evolve to the state at the end of the measurement

$$\psi(x, X) = \varphi_0(x)\Phi_0(X - gx), \tag{2}$$

where

$$g = \frac{1}{T} \int_0^T f(t) dt \tag{3}$$

is a dimensionless coupling constant. The final state (2) of the system is the product of the state of the particle, unaffected by the measurement, and the state of the pointer indicating the result of the measurement. For a measurement that is very fast the wavepacket for the pointer has not spread, and its center has been displaced by an amount given by the shift $s(x) = gx$. Furthermore, the probability distribution that in the final state the pointer indicates a value corresponding to the position $x$ of the particle is

$$P(x) = \int_{-\infty}^{\infty} |\varphi_0(x)\Phi_0(X - gx)|^2 \, dX. \tag{4}$$

For a normalized initial pointer state $\Phi_0(X)$ this is

$$P(x) = |\varphi_0(x)|^2. \tag{5}$$

This is in agreement with the general characterization of measurement theory in reference [2]: The probability distribution in the final state that the apparatus indicates the value $x$, is equal to $|\varphi_0(x)|^2$, just the same as the probability distribution for the position of the particle in the state before the measurement. It is in this sense that this is a perfect measurement.

In this paper, we will consider the measurement of the position of the particle during a finite time $T$ comparable to dynamical times for the pointer and the particle, which is more realistic than considering an instantaneous measurement. Thus the motion of the particle and the pointer cannot be neglected. It must be noted that the present work is not a continuous measurement of the position in the sense of references [3] and [4]. It is simply a single measurement that takes a finite time to be performed. A similar study was done in reference [5] for the spin of a particle. References [6] – [8] also develop a formalism for measuring the time average of any dynamical quantity on individual Feynman histories; and a Feynman meter, implicitly with infinite mass, was introduced. In particular, the time average of the position of a particle is discussed in reference [6]. In the present work we are not interested in calculating averages on particular Feynman histories. Rather we will consider the unitary evolution of the system particle-pointer by summing over all Feynman histories in order to obtain the wavefunction at the end of the measurement. Specifically we are interested in what the pointer will indicate at the end of the finite time measurement and what can be said in this case about probability distributions for the position of the particle. We will find that for a particle with a sharp value of its initial position, at the end of the measurement the pointer will indicate the arithmetic average between the initial and final positions; in addition, the probability that the pointer will indicate a given average value is equal to the transition probability for the undisturbed particle to experience the change in position.

We will consider a system of two particles: one of mass $m$ whose position we want to measure, and the other is the pointer of mass $M$. The momentum operators for the particle and the pointer are $p$ and $P$ respectively. The position of the particle and the indication of the pointer will be correlated via the interaction (1) and the Hamiltonian of this system is

$$H = H_0 + H_i \tag{6}$$

with

$$H_0 = \frac{p^2}{2m} + V(x) + \frac{P^2}{2M}, \qquad (7)$$

where $H_i$ is given by (1) and $V(x)$ is the potential acting on the particle. We will restrict ourselves to two tractable cases for $V(x)$: i) a free particle $(V = 0)$ with any masses for particle and pointer. The coupling function will be a constant during the measurement. ii) A simple harmonic oscillator with a pointer of infinite mass. Exact solutions for the time evolution of the wavefunction of these systems will be obtained. Now because the kinetic energy of the particle cannot be ignored in a finite time measurement, the position of the particle is no longer a constant of the motion, unlike the case with (1) alone, and a perfect measurement is no longer possible [9], [10]. That is, the distribution of the position of the particle in the state just before the measurement cannot be obtained from the state of the system when the measurement has been completed.

The current problem is also of interest in the generalized quantum mechanics of closed systems [11] and [12]. For macroscopic masses, in particular, there is a class of spacetime coarse grainings which extend over time to which nearly instantaneous alternatives yield an excellent approximation with respect to decoherence and to probabilities [13]. In realistic measurement situations of light systems which extend over time, the measuring apparatus should be taken into account, so that the different outcomes of the measurement should correlate with different readings of the pointer. Even though a perfect measurement is no longer possible, if the motion of the particle cannot be neglected, it is still of interest to treat the particle and apparatus as a dynamical system and to follow the motion of the pointer as an indicator of the position of the particle, both when the measurement is completed, and also while the measurement is taking place.

The model described by (6) and (7) is developed in section 2 of this paper, and its behavior for a pointer with infinite mass is found in section 3. In section 4 we examine the cases of finite masses for the pointer and the particle, and infinite mass for the pointer. Different solutions are considered, and their significance is discussed. In section 5 we study the probability distribution for the position of the particle when the measurement is completed.

## 2. Formalism. Application to the Free Particle Case

In this section we will outline how to construct the propagator for the Hamiltonian (6). The fine grained histories for this system are the particle paths $x(t)$ and the pointer paths $X(t)$ on a time interval $[0,T]$. We begin with the discrete case of partitioning the interval into $N$ subintervals $[t_n, t_{n-1}]$ with

$$t_n = n\frac{T}{N} \qquad n = 0, 1, \ldots, N \tag{8}$$

In the limit $N \to \infty$ the fine grained histories will be recovered.

Next we introduce the interaction Hamiltonian in the Schrödinger picture

$$H_i = xP\frac{1}{N}\sum_{n=0}^{N-1} f(t_n)\delta(t - t_n). \tag{9}$$

That is, as $N$ gets larger, the measurement process can be approximated better and better as a sequence of $N$ instantaneous von Neumann measurements. In between measurements the particle and pointer are coupled and evolve according to (7). Letting $N \to \infty$ we obtain the limiting case of a von Neumann measurement of finite duration.

Regarding time evolution, the quantity of interest is $\int_0^T H_i dt$. For the case of (9) this yields

$$\int_0^T H_i dt = xP\frac{1}{T}\sum_{n=0}^{N-1} \varepsilon f(t_n) \tag{10}$$

where

$$\varepsilon = t_{n-1} - t_n = \frac{T}{N} \tag{11}$$

and

$$\int_0^T H_i(t) \mathrm{d}t \underset{\varepsilon \to 0}{=} xP \frac{1}{N} \int_0^T f(t) \mathrm{d}t . \tag{12}$$

This is the same as $\int_0^T H_i \mathrm{d}t$ obtained using $H_i$ given by (1) and the constant $g$ given by (3). Thus in the limiting case $N \to \infty$, the time evolution due to (1) is equivalent to that due to (9). The factor $1/N$ in (9) ensures that as $N$ increases, the strength of the Dirac delta functions gets weaker. In the limit of very large $N$ we get the same time evolution as with (1).

The time evolution operator can be written

$$U(T,0) = U_0(t_N, t_{N-1}) \cdots U(t_n) U_0(t_n, t_{n-1}) U(t_{n-1}) \cdots U(t_1) U_0(t_1, t_0) U(t_0) \tag{13}$$

where the $U(t_n)$ factors are the time evolution operators corresponding to instantaneous von Neumann measurements. From here on we set the value of the reduced Planck constant equal to one.

$$U(t_n) \underset{\delta \to 0}{=} \exp\left[-i \int_{t_n - \delta}^{t_n + \delta} H_i(t) \mathrm{d}t\right] = \exp\left[-\frac{i}{N} f(t_n) xP\right]. \tag{14}$$

$U_0(t_n, t_{n-1})$ is the time evolution operator for the uncoupled particle and pointer:

$$U_0(t_n, t_{n-1}) = \exp\left[-iH_0(t_n - t_{n-1})\right]. \tag{15}$$

Next we form the matrix element $\langle x_N, X_N | U(T,0) | x_0, X_0 \rangle$, with $U(T,0)$ given by (13). Introducing $N-1$ complete sets of position states for the particle and for the pointer and noting that

$$U(t_n) | x_n, X_n \rangle = \left| x_n, X_n + \frac{1}{N} f(t_n) x_n \right\rangle \tag{16}$$

we can rewrite

$$\langle x_N, X_N | U(T,0) | x_0, X_0 \rangle =$$

$$\int_{-\infty}^{\infty} dx_{N-1} \int_{-\infty}^{\infty} dX_{N-1} \langle x_N, X_N | U_0(t_N, t_{N-1}) | x_{N-1}, X_{N-1} + \frac{1}{N} f(t_{N-1}) x_{N-1} \rangle$$

$$\cdots \int_{-\infty}^{\infty} dx_n \int_{-\infty}^{\infty} dX_n \langle x_n, X_n | U_0(t_n, t_{n-1}) | x_{n-1}, X_{n-1} + \frac{1}{N} f(t_{n-1}) x_{n-1} \rangle \quad (17)$$

$$\cdots \int_{-\infty}^{\infty} dx_1 \int_{-\infty}^{\infty} dX_1 \langle x_1, X_1 | U_0(t_1, t_0) | x_0, X_0 + \frac{1}{N} f(t_0) x_0 \rangle.$$

For small $\varepsilon$ we have the matrix element:

$$\langle x_n, X_n | U_0(t_n, t_{n-1}) | x_{n-1}, X_{n-1} + \frac{1}{N} f(t_{n-1}) x_{n-1} \rangle =$$

$$\left(\frac{\lambda_m}{i\pi}\right)^{1/2} e^{i[\lambda_m (x_n - x_{n-1})^2 - \varepsilon V(x_{n-1})]} \left(\frac{\lambda_M}{i\pi}\right)^{1/2} e^{i\lambda_M \left[X_n - X_{n-1} - \frac{1}{N} f(t_{n-1}) x_{n-1}\right]^2}, \quad (18)$$

where the quantities $\lambda_m$ and $\lambda_M$ are given by

$$\lambda_m = \frac{m}{2\varepsilon} \quad (19)$$

and

$$\lambda_M = \frac{M}{2\varepsilon}, \quad (20)$$

with $\varepsilon$ given by (11). Letting $x_0 = x'$, $X_0 = X'$; $x_N = x''$, $X_N = X''$ and inserting (18) into (17) we take the limit $N \to \infty$, to obtain the path integral expression for the matrix element $\langle x'', X'' | U(T,0) | x', X' \rangle$ extended to all paths starting at $(x', X')$ and ending at $(x'', X'')$:

$$\langle x'', X'' | U(T,0) | x', X' \rangle = \iint \delta x \delta X e^{iS[x(t), X(t)]}, \quad (21)$$

where the action is

$$S[x(t), X(t)] = \int_0^T \left[ \frac{m}{2} \dot{x}^2 - V(x) + \frac{M}{2}\left(\dot{X} - \frac{f(t)}{T}x\right)^2 \right] dt. \quad (22)$$

Note also, that the Hamiltonian (6) easily follows from this action.

Expression (13) for the time evolution operator of the system suggests the following factorization for the path integral in (21):

$$\langle x'', X'' | U(T,0) | x', X' \rangle = \langle x'' | U_{particle}(T,0) | x' \rangle G(X'', X'; x'', x'; T) \quad (23)$$

where

$$U_{particle}(T,0) = \exp\left[-i\left(\frac{p^2}{2m} + V(x)\right)T\right] \quad (24)$$

is the time evolution operator for an undisturbed particle and $V(x)$ is smooth. The factor $G(X'', X'; x'', x'; T)$ in (23) is associated with the pointer and is of the form

$$G(X'', X'; x'', x'; T) = \left(\frac{M_{eff}(T)}{2\pi i T}\right)^{1/2} \exp\left\{i \frac{M_{eff}(T)}{2T}[X'' - X' - s(x'', x', T)]^2\right\}. \quad (25)$$

Expression (25) is the propagator for a free particle which starts at the shifted position $X' + s(x'', x', T)$ and ends at $X''$ at time $T$. The effective mass $M_{eff}$ depends on the masses of the pointer and the particle as well as $T$. The shift function $s(x'', x', T)$ depends on the initial and final position of the particle and the duration of the measurement. The form of the effective mass and the shift function at time $T$ will depend on the particle potential $V(x)$ and the time dependent coupling function $f(t)$.

The conjecture expressed by (23) is motivated by the observation that, starting with the particle at $x'$ and the pointer at $X'$, at each of the instantaneous measurements (14) the particle part of the state of the system is unchanged by the measurement, as shown by (2). Also, from one instantaneous measurement of the position to the next, the particle part of the state evolves according to (24), and the pointer part evolves freely.

This holds true for any value of $N$, and so that it can be true in the limit of very large $N$. Thus pointer and particle evolve independently of each other, except when the instantaneous measurements take place where they couple together. The effect of each instantaneous successive measurement is to add a shift to the position of the pointer, which then proceeds to evolve freely until the next measurement takes place. After taking the limit $N \to \infty$, the result is that the pointer has propagated like a free particle with an effective mass and its initial position has experienced a shift given by the shift function $s(x'', x', T)$. We will verify this for the case of a free particle, with any masses for the pointer and the particle, and also for the simple harmonic oscillator when the pointer has infinite mass. In both instances we will find that the overall shift recorded by the pointer is proportional to the arithmetic average for the initial and final position of the particle provided the coupling function in (1) is symmetric about the midpoint of the interval $[0,T]$.

Next, for the sake of simplicity, we set coupling function $f(t) = g$, a constant in the interval $[0,T]$, and evaluate the path integral for this system for the case of a free particle. Since the action (22) is quadratic, the semiclassical approximation becomes exact [14]:

$$\iint \delta x \delta X e^{iS[x(t),X(t)]} = A e^{iS_{cl}} \qquad (26)$$

where $S_{cl}$ is the action evaluated at the classical path $x_{cl}(t)$, $X_{cl}(t)$ and $A$ is a normalization constant.

The action (22) yields the classical equation of motion

$$m\ddot{x} + \frac{gM}{T}\dot{X} - \frac{g^2 M}{T^2} x = 0 \qquad (27)$$

plus the conservation of the canonical momentum of the pointer

$$\frac{dP}{dt} = \frac{d}{dt}\left[ M\left(\dot{X} - \frac{gx}{T}\right)\right] = 0. \qquad (28)$$

Letting $x' = x(0)$, $X' = X(0)$ and $x'' = x(T)$, $X'' = X(T)$, equations (27) and (28) can be Solved exactly:

$$x_{cl}(t) = x' + \left(x'' - x' + \kappa X_{0T}\right)\frac{t}{T} - \kappa X_{0T}\frac{t^2}{T^2} \tag{29}$$

$$\dot{X}_{cl}(t) = \frac{g}{T}x_{cl}(t) + \frac{1}{T}X_{0T} \tag{30}$$

with

$$\kappa = \frac{gM}{2m} \tag{31}$$

and

$$X_{0T} = \frac{X'' - X' - g(x'' + x')/2}{1 + g\kappa/6}. \tag{32}$$

The classical action for this system is

$$S_{cl} = S\left[x_{cl}(t), X_{cl}(t)\right] = \frac{1}{T}\left\{\frac{m}{2}(x'' - x')^2 + \frac{M_{eff}}{2}\left[X'' - X' - g(x'' + x')/2\right]^2\right\}, \tag{33}$$

where

$$M_{eff} = \frac{M}{1 + g\kappa/6} \tag{34}$$

is a "renormalized" or effective mass. The normalization constant in (26) can be obtained from application of the Van Vleck-Pauli-Morette determinant [15], which for the classical action (33) yields $A = \frac{\sqrt{mM_{eff}}}{2\pi iT}$.

Finally, the exact propagator for this system is

$$\langle x'', X''|U(T,0)|x', X'\rangle = \frac{\sqrt{mM_{eff}}}{2\pi iT}\exp\left\{i\frac{m}{2T}(x''-x')^2 \right.$$
$$\left.+i\frac{M_{eff}}{2T}\left[X''-X'-g\frac{x''+x'}{2}\right]^2\right\}. \quad (35)$$

Expression (35) is a product of two factors of the form anticipated in (23): The first factor is the particle part, which in this case is the free propagator

$$\langle x''|U_{particle}(T,0)|x', X'\rangle = \sqrt{\frac{m}{2\pi iT}}\exp\left[i\frac{m}{2T}(x''-x')^2\right] \quad (36)$$

and the second factor is a portion associated with the pointer

$$G(X'', X'; x'', x'; T) = \left(\frac{M_{eff}}{2\pi iT}\right)^{1/2}\exp\left\{i\frac{M_{eff}}{2T}\left[X''-X'-g\frac{x''+x'}{2}\right]^2\right\}. \quad (37)$$

The expression above is the propagator for a free particle of mass $M_{eff}$, where the initial position of the pointer has been shifted by an amount proportional to the arithmetic average of the initial and final position of the particle. Also the shift in position is independent of the masses

$$s(x'', x', T) = g\frac{x''+x'}{2}. \quad (38)$$

As noted earlier, the pointer propagates like a free particle between the instantaneous measurements. The net effect of taking the limit $N \to \infty$ is to give an effective mass (34) to the pointer and a shift in its position given by (38).

As in the case of (35), at any time $t < T$, while the measurement is taking place, the semiclassical approximation is exact. The result is that the propagator for the system is obtained by replacing the coupling constant $g$ in (31) and in (35) by $g(t) = g\frac{t}{T}$.

Equation (35) provides a dynamical description for the finite time measurement of the position when the particle and the pointer have finite masses. In this case their motion must be taken into account. In addition, one can get the state of the system when the measurement has been completed as well as while it is taking place. This will be done in section 4.

## 3. Pointer with Infinite Mass

It is instructive to consider the case that the pointer has infinite mass. Pointers with infinite mass are also considered in references [5] – [8]. The kinetic term for the pointer in (7) is absent, and the pointer responds to the interacting term (1) only. That is, the kinetic motion of the pointer may not be so essential after all, since we are primarily interested in the basic features of the indicator function of the apparatus during and right after the measurement. In particular, this will make it easier to analyze the case of the simple harmonic oscillator, which will be developed in this section. Also, as will be shown, pointers with infinite mass provide a natural way to record the time average

$$\bar{x}[x(t)] = \frac{1}{T}\int_0^T x(t) \mathrm{d}t \qquad (39)$$

of the position of the particle on individual histories.

The pointer is now at rest at the origin and it simply shifts its position in response to (1), recording the successive positions of the particle as it moves. The time interval is partitioned into $N$ steps defined by (8), as was done in the previous section, and the Hamiltonian for the system with a constant coupling function $g$ in (1) is

$$H = \frac{p^2}{2m} + V(x) + gxP\frac{1}{N}\sum_{n=0}^{N-1}\delta(t-t_n) \ . \qquad (40)$$

Starting with the sharp position state $|x_0\rangle|X_0\rangle$ for the particle and the pointer, and introducing complete sets of states for the particle, unitary evolution yields the state of the system at time $T$ :

$$U(T,0)|x_0\rangle|X_0\rangle = \int_{-\infty}^{\infty} dx_{N-1} U_0(t_N,t_{N-1})|x_{N-1}\rangle$$

$$\cdots \int_{-\infty}^{\infty} dx_n \langle x_n|U_0(t_n,t_{n-1})|x_{n-1}\rangle \quad (41)$$

$$\cdots \int_{-\infty}^{\infty} dx_1 \langle x_1|U_0(t_1,t_0)|x_0\rangle \left| X_0 + g\frac{1}{N}\sum_{n=0}^{N-1} x_n \right\rangle$$

where

$$U_0(t_n,t_{n-1}) = \exp\left\{-i\left(\frac{p^2}{2m} + V(x)\right)(t_n - t_{n-1})\right\}. \quad (42)$$

That is, for any particular particle path, the pointer has experienced an overall shift in its position given by an amount proportional to the average of all the intermediate positions of the particle along a path. In the limit $N \to \infty$ the time average (39) is obtained.

In expression (41) for the time evolution of the initial position state, we have the matrix elements:

$$\langle x_n|U_0(t_n,t_{n-1})|x_{n-1}\rangle = \sqrt{\frac{\lambda_m}{i\pi}} \exp\left\{i\left[\lambda_m(x_n - x_{n-1})^2 - \varepsilon V(x_{n-1})\right]\right\} \quad (43)$$

with $\varepsilon$ given by (11) and $\lambda_m$ is given by (19).

Putting $x_0 = x'$, $X_0 = X'$ and substituting (43) into (41), we multiply from the left by the state $\langle x'', X''|$ and take the limit $N \to \infty$, which yields the path integral:

$$\langle x'', X''|U(T,0)|x', X'\rangle = \int \delta x\, e^{i\int_0^T \left[\frac{m}{2}\dot{x}^2 - V(x)\right]dt} \delta(X'' - X' - g\bar{x}) \quad (44)$$

with $\bar{x} = \bar{x}[x(t)]$ the time average over a particular path for a particle that that starts at $x'$ and ends at $x''$. On an individual Feynman path the pointer position has shifted by an amount proportional to the time average of the particle's position on the path.

Introducing the Fourier representation for the Dirac delta function

$$\delta \left( X'' - X' - g\bar{x} \right) = \frac{1}{2\pi} \int_{-\infty}^{\infty} dP\, e^{iP(X'' - X' - g\bar{x})} \qquad (45)$$

and after substituting (45) into (44), and inserting $\bar{x}$ given by (39), we get the matrix element

$$\langle x'', X'' | U(T,0) | x', X' \rangle = \frac{1}{2\pi} \int_{-\infty}^{\infty} dP\, e^{iP(X'' - X')} \int \delta x\, e^{iS_{\mathit{eff}}[x(t)]} , \qquad (46)$$

where

$$S_{\mathit{eff}}[x(t)] = \int_0^T \left[ \frac{m}{2} \dot{x}^2 - V_{\mathit{eff}}(x) \right] dt , \qquad (47)$$

is an effective action, and

$$V_{\mathit{eff}}(x) = \frac{g}{T} xP + V(x) \qquad (48)$$

is and effective potential, which includes the potential $V(x)$ acting on the particle. If we replace the constant $g$ by the coupling function $f(t)$ in the Hamiltonian (40), then the expression (48) above for the effective potential will exhibit $f(t)$ instead of the coupling constant $g$.

Next we will consider the two cases: a) a free particle with a constant coupling function $f(t)$, and (b) a harmonic oscillator.

**a) Free Particle**

The path integral for the free particle is

$$I_{\mathit{free}} = \int \delta x \exp\left[ i \int_0^T \left( \frac{m}{2} \dot{x}^2 - \frac{g}{T} xP \right) dt \right] \qquad (49)$$

which is a well known result [14]:

$$I_{free} = \sqrt{\frac{m}{2\pi iT}} \exp\left\{i\left[\frac{m}{2T}(x''-x')^2 - gP\left(\frac{x''+x'}{2}\right) - \frac{g^2 P^2 T^2}{24m}\right]\right\}. \qquad (50)$$

Substituting (50) into (46) and evaluating the Fourier transform we obtain the result for the propagator

$$\langle x'', X'' | U(T,0) | x', X' \rangle = \sqrt{\frac{m}{2\pi iT}} \sqrt{\frac{6m}{i\pi g^2 T}} \cdot$$

$$\exp\left\{\frac{im}{2T}(x''-x')^2 + \frac{6im}{g^2 T}\left[X'' - X' - g\left(\frac{x''+x'}{2}\right)\right]^2\right\} \qquad (51)$$

which agrees with letting $M \to \infty$ in the finite mass propagator (35).

**b) Harmonic Oscillator**

The particle is a harmonic oscillator of frequency $\omega$ with potential $V(x) = \frac{1}{2}m\omega^2 x^2$. Then the path integral that we need to consider in this case is

$$I_{SHO} = \int \delta x \exp\left[i\int_0^T \left(\frac{m}{2}\dot{x}^2 - \frac{m}{2}\omega^2 x^2 - \frac{f(t)}{T}xP\right)dt\right] \qquad (52)$$

where we have included the coupling function from (1). This is another well-known result [14]:

$$I_{SHO} = \sqrt{\frac{m\omega}{2\pi i \sin \omega T}} \exp\left\{ i \frac{m\omega}{2 \sin \omega T} \left[ \left( x''^2 + x'^2 \right) \cos \omega T - 2x''x' \right.\right.$$

$$\left. - \frac{2x''P}{m\omega T} \int_0^T dt\, f(t) \sin \omega t - \frac{2x'P}{m\omega T} \int_0^T dt\, f(t) \sin \omega (T-t) \right. \tag{53}$$

$$\left.\left. - \frac{2P^2}{m^2\omega^2 T^2} \int_0^T dt \int_0^t dt'\, f(t) f(t') \sin \omega (T-t) \sin \omega t' \right] \right\}.$$

If we assume that the coupling function $f(t)$ is symmetrical about the midpoint of the time interval $[0,T]$, then the two integrals in the linear term in the pointer momentum in (53) are equal and the expression (46) for the propagator of the system becomes

$$\langle x'', X'' | U(T,0) | x', X' \rangle = K_{SHO}(x'', x', T) \cdot \tag{54}$$

$$\frac{1}{2\pi} \int_0^T dP\, e^{iP(X''-X')} \exp\left\{ -i \frac{2P}{T \sin \omega T} B(\omega, T) \left( \frac{x'' + x'}{2} \right) - i \frac{P^2}{m\omega T^2 \sin \omega T} A(\omega, T) \right\},$$

where

$$K_{SHO}(x'', x', T) = \sqrt{\frac{m\omega}{2\pi i \sin \omega T}} \exp\left\{ i \frac{m\omega}{2 \sin \omega T} \left[ \left( x''^2 + x'^2 \right) \cos \omega T - 2x''x' \right] \right\} \tag{55}$$

is the expression for the propagator for the harmonic oscillator when the oscillator is not coupled to the pointer. The two functions $A(\omega, T)$ and $B(\omega, T)$ are given by:

$$A(\omega, T) = \int_0^T dt \int_0^t dt'\, f(t) f(t') \sin \omega (T-t) \sin \omega t' \tag{56}$$

$$B(\omega, T) = \int_0^T dt\, f(t) \sin \omega t \tag{57}$$

After doing the integration over the momentum $P$, we get the propagator of this system:

$$\langle x'', X'' | U(T,0) | x', X' \rangle = K_{SHO}(x'', x', T) \cdot$$

$$\sqrt{\frac{M_{eff}(T)}{2\pi i T}} \exp\left\{ i \frac{M_{eff}(T)}{2T} \left[ X'' - X' - g(T) \frac{x'' + x'}{2} \right]^2 \right\}$$
(58)

with

$$M_{eff}(T) = \frac{m\omega T^3 \sin \omega T}{2A(\omega, T)}$$
(59)

and

$$g(T) = \frac{2B(\omega, T)}{T \sin \omega T}.$$
(60)

Again result (58) is of the form anticipated in (23). The particle propagates in this case like a simple harmonic oscillator, unaffected by the measurement process, and the pointer propagates like a free particle with an effective mass given by (59). The pointer experiences and overall shift in position independent of the mass of the particle:

$$s(x'', x', T) = g(T) \frac{x'' + x'}{2},$$
(61)

where the coupling constant $g(T)$ is given by (60).

In the general case that the coupling is not symmetrical about $T/2$ we get a shift function which will not be proportional to the average of $x'$ and $x''$. Still, in this case, the displacement of the pointer will be a linear combination of the initial and final position of the particle.

If we let the coupling function $f(t) = g$, a constant, then expression (59) for the effective mass reduces to

$$M_{eff}(T) = m \frac{\omega^2 T^2}{g^2 \left[ \frac{\tan(\omega T/2)}{\omega T/2} - 1 \right]}$$
(62)

and the coupling constant (60) becomes

$$g(T) = g\frac{\tan(\omega T/2)}{\omega T/2} \quad . \tag{63}$$

We readily obtain the result the result for the free particle propagator, when the coupling function in the interaction (1) is symmetrical about the midpoint $T/2$, by taking the limit $\omega \to 0$ in (58):

$$\langle x'',X''|U(T,0)|x',X'\rangle \underset{\omega \to 0}{=} \sqrt{\frac{m}{2\pi iT}} \exp\left[i\frac{m}{2T}(x''-x')^2\right] \cdot$$

$$\sqrt{\frac{M_{\mathit{eff}}(T)}{2\pi iT}} \exp\left\{i\frac{M_{\mathit{eff}}(T)}{2T}\left[X''-X'-g(T)\frac{x'+x''}{2}\right]^2\right\} \tag{64}$$

From (56) the effective mass (59) becomes

$$M_{\mathit{eff}}(T) \underset{\omega \to 0}{=} mT^4 \Big/ \left[2\int_0^T dt \int_0^t dt'\,(T-t)t'f(t)f(t')\right] \tag{65}$$

and from (60) the coupling constant becomes

$$g(T) = \frac{2}{T^2}\int_0^T t\,f(t)dt \quad . \tag{66}$$

For particle potentials of the form $V(x) = V_0 + V_1 x + V_2 x^2$, where $V_0, V_1, V_2$ are constants, the semiclassical approximation (26) is exact, and the shift in position will now be a linear function of the initial and final position of the particle instead of their average: $s(x'',x',T) = a(T)x'' + b(T)x' + c(T)$ even with the coupling function being symmetrical about the midpoint $T/2$. This is due to the presence of the linear term in $x$ in the potential.

Finally, with a constant coupling function, the previous analysis for the harmonic oscillator yields the result for the propagator of the system while the measurement is taking place:

$$\langle x'', X'' | U(t,0) | x', X' \rangle = K_{SHO}(x'',',t) \cdot$$

$$\sqrt{\frac{M_{eff}(t)}{2\pi i t}} \exp\left\{ i \frac{M_{eff}(t)}{2t} \left[ X'' - X' - g(t) \frac{x'' + x'}{2} \right]^2 \right\}, \tag{67}$$

where now the effective mass of the pointer is

$$M_{eff}(t) = m \frac{\omega^2 T^2}{g^2 \left[ \frac{\tan(\omega t/2)}{\omega t/2} - 1 \right]} \tag{68}$$

with $t < T$. This is a monotonically decreasing function of $t$ which is infinite at the start of the measurement, as it should be, and decreases to its final value (62) at the end of the measurement. The coupling constant in (67) is

$$g(t) = \frac{gt}{T} \frac{\tan(\omega t/2)}{\omega t/2} \tag{69}$$

which increases from zero at the start of the measurement to its final value given by (63) at the end of the measurement. The value of the coupling constant is infinite at $t = \pi/\omega$; thus the duration of the measurement can be any time $T$ such that $T < \pi/\omega$.

To conclude this section, pointers of infinite mass are worth considering when our primary interest is the basic features of the indicator or shift function of the apparatus during and at the end of the measurement, uncluttered by the kinetic motion of the pointer. In realistic measurement situations, pointers are much more massive than the particle whose position we want to measure; thus it makes sense to consider this case. Also, letting the mass of the pointer become infinite makes the study of quadratic and linear particle potentials much easier to study. At the same time we have been able to consider the effect that the coupling function in the interaction (1) has on the propagator of the system, both for the harmonic oscillator and for a free particle. In both cases, when the coupling function is symmetrical about the midpoint of the interval $[0,T]$, the pointer indicates the average value of the initial and final position of the particle. In the next section we will apply the results obtained in sections 2 and 3 to investigate the form of the wave function of the system at the end of the measurement.

## 4. State of the System During and after the Measurement

Unitary evolution will determine the state of the system particle-pointer at any time $t \leq T$. That is,

$$\psi(x,X,t) = \int_{-\infty}^{\infty}\int_{-\infty}^{\infty} dx'dX' \langle x,X|U(t,0)|x',X'\rangle \psi_0(x',X') \qquad (70)$$

with the initial state of the system $\psi_0(x,X) = \varphi_0(x)\Phi_0(X)$, where $\varphi_0(x)$ and $\Phi_0(X)$ are the initial states of the particle and the pointer respectively.

If we take the limit $T \to 0$ in (35) for the free particle or in (58) for the harmonic oscillator and insert into (70) we recover the characteristic von Neumann result (2). At any time during the measurement $t < T$, the limit $t \to 0$ yields the initial state $\psi_0(x,X)$ before the measurement.

Next we will insert specific initial states into (70). We will consider the two cases both for the free particle and for the harmonic oscillator:

I. The initial state of the particle has a sharp value of the position and the pointer is described by a wavepacket $\Phi_0(X)$ centered at $X = 0$: $\psi_0(x,X) = \delta(x - x_0)\Phi_0(X)$

For the free particle, inserting this initial state and the propagator (35) into (70) yields the state at the end of the measurement:

$$\psi(x,X,T) = \varphi_m(x - x_0,T)\Phi_{M_{eff}}(X - g\bar{x},T) \qquad (71)$$

with

$$\varphi_m(x - x_0, T) = \sqrt{\frac{m}{2\pi iT}} e^{i\frac{m}{2T}(x-x_0)^2} \tag{72}$$

and

$$\Phi_{M_{eff}}(X - g\bar{x}, T) = \sqrt{\frac{M_{eff}}{2\pi iT}} \int_{-\infty}^{\infty} dX' \Phi_0(X') \exp\left[i\frac{M_{eff}}{2T}(X - X' - g\bar{x})^2\right], \tag{73}$$

where $M_{eff}$ is the effective mass given by (34) and $\bar{x}$ is the arithmetic mean between values of the position in the initial state and the position in the final estate.

The result (71) for the final state shows the von Neumann entanglement corresponding to undisturbed free evolution for the particle part of the wavefunction. The wavepacket (73) describing the pointer has spread and shifted, indicating the average position of the particle over time. It can be rewritten in the form

$$\Phi_{M_{eff}}(X - g\bar{x}, T) = \sqrt{\frac{M_{eff}}{2\pi iT}} \int_{-\infty}^{\infty} dX' \Phi_0(X' - g\bar{x}) \exp\left[i\frac{M_{eff}}{2T}(X - X')^2\right]. \tag{74}$$

This is the expression for a spreading wavepacket centered at $X_0 = g\bar{x}$ at $t = 0$. At time $T$ the pointer has the spread corresponding to the evolution as a free particle with mass $M_{eff}$.

II. The initial state of the particle does not have a sharp value of the position, and the pointer is described by a wavepacket $\Phi_0(X)$ centered at $X = 0$: $\psi_0(x, X) = \varphi_0(x)\Phi_0(X)$.

Now for the free particle expression (70) yields

$$\psi(x, X, T) = \int_{-\infty}^{\infty} dx' \varphi_0(x') \varphi_m(x - x', T) \Phi_{M_{eff}}(X - g\bar{x}, T) \tag{75}$$

with $\bar{x} = (x' + x)/2$ and $x_0$ is replaced by $x'$ in the expression (72) describing a free particle, and in (73) describing the pointer. This is a superposition of states of the form (71) obtained in the previous case. In the limit of an instantaneous measurement (75) yields the familiar von Neumann result (2).

According to the result (75), if the range of $x'$ in the initial state $\varphi_0(x')$ of the particle is $-\Delta < x < \Delta$, then after an observation of the system when the measurement is completed, for a given final position $x$ of the particle the pointer has moved, and its possible final positions $\bar{x}$ spread continuously between $-\Delta/2$ and $\Delta/2$.

Guided by the previous case (I) above, we could consider a free particle whose initial state is represented by a very narrow wavepacket centered at $x_1$ and with spread $\delta(0)$. If the particle evolved without any interaction with the pointer, then at time $T$ the wavepacket would have traveled and would be centered at $x_2$ with the spread at this time being small, $\delta(T)$, and with $\delta(0) < \delta(T)$. That is, the time evolution for this free particle wavepacket would be

$$\varphi(x,T) = \sqrt{\frac{m}{2\pi iT}} \int_{-\infty}^{\infty} dx' \varphi_0(x') e^{i\frac{m}{2T}(x-x')^2} , \qquad (76)$$

where the main contribution to the integral comes from values of $x'$ in the interval $x_1 - \delta(0) \leq x \leq x_1 + \delta(0)$. If we assume that during this time the wavepacket has remained narrow, then when the particle couples with the pointer, the right hand side in expression (75) can be approximated by

$$\psi(x,T) \approx \left( \int_{-\infty}^{\infty} dx' \varphi_0(x') \varphi_m(x - x', T) \right) \Phi_{M_{eff}} \left( X - g\bar{x}, T \right) \qquad (77)$$

and at the end of the measurement the pointer is centered within the range

$$\frac{x_1 + x_2}{2} - \Delta \leq \bar{x} \leq \frac{x_1 + x_2}{2} + \Delta \qquad (78)$$

where the spread of $\bar{x}$ is

$$\Delta = \frac{\delta(0) + \delta(T)}{2} \qquad (79)$$

The velocity of the wavepacket is $v = \frac{x_2 - x_1}{T}$ and in the initial state $\psi_0(x, X) = \varphi_0(x)\Phi_0(X)$, we let $\varphi_0(x) = A(x - x_1)e^{ip_0(x - x_1)}$ with momentum $p_0 = mv$. For a Gaussian wavepacket

$$A(x - x_1) = \left(\frac{2}{\pi l^2}\right)^{1/4} e^{-(x - x_1)^2/l^2} \qquad (80)$$

the spread is given by

$$\delta(t) = l\left(1 + \frac{4t^2}{m^2 l^4}\right)^{1/2}. \qquad (81)$$

We require

$$\frac{4T^2}{m^2 l^4} \ll 1 \qquad (82)$$

so that $\delta(T) \approx \delta(0)$ and the wavepacket has not spread too much. Under these conditions the approximation leading to expression (77) may be justified, and from (79) and (81) we obtain the result

$$\Delta \approx l + \frac{T^2}{m^2 l^3}, \qquad (83)$$

where from (82), the duration of the measurement is such that

$$T \ll \frac{ml^2}{2}. \qquad (84)$$

For example, for an electron within an initial spread of $0.1\overset{0}{\text{A}}$, $T$ is very short, less than $10^{-20}$ sec. For a speck of dust with $m = 10^{-6}$ kg and spread $10^{-6}$ m, $T$ can be very long, up to $10^8$ years.

In the approximate expression (77), the factor representing the pointer can be rewritten as in (74). Thus for a pointer whose initial state is a Gaussian centered at the origin

$$\Phi_0(X) = \left(\frac{2}{\pi L^2}\right)^{1/4} e^{-X^2/L^2} \tag{85}$$

expression (81) for the spread shows that the measurement must be completed in time $T$ before the spread of the pointer doubles in size. That is, $T < T_{double}$, where

$$T_{double} = \frac{\sqrt{3}}{2} M_{eff} L^2 . \tag{86}$$

If we further require that the pointer does not double in size within the time the measurement lasts, then from (84) and (86), the effective mass of the pointer is such that

$$M_{eff} > \frac{ml^2}{\sqrt{3}L^2} . \tag{87}$$

As noted earlier, expression (74) for the pointer describes a spreading wavepacket centered at $\bar{x}$. As time goes on, the width $\Delta(T)$ of the packed increase and eventually becomes so broad that, for instance, for an initial pointer state given by a Gaussian centered at the origin we can write

$$\Phi_{M_{eff}}(X - g\bar{x}, T) \approx \Phi_{M_{eff}}(X, T) + O(1/\Delta(T)) . \tag{88}$$

In this case (75) can be approximated by

$$\psi(x, X, T) \approx \varphi(x, T) \Phi_{M_{eff}}(X, T), \tag{89}$$

where $\varphi(x,T)$ is the time evolution (76) of the free particle state. That is, the pointer is centered at $X = 0$ and does not provide any information about the average position of the particle.

When the particle is acted on by a potential $V(x)$, for instance the harmonic potential, expression (75) for the free particle is replaced by

$$\psi(x,X,T) = \int_{-\infty}^{\infty} \varphi_0(x')\varphi_{particle}(x,x',T)\Phi_{M_{eff}}\left(X - s(x,x',T)\right) dx' \qquad (90)$$

where the second factor in the integrand is now the particle propagator $\langle x|U_{particle}(T,0)|x'\rangle$, with $U_{particle}$ given by (24). The third factor in the integrand is obtained from expression (74) for the pointer. The effective mass and the shift function depend on the form of the potential acting on the particle. For the simple harmonic oscillator, when the coupling function $f(t)$ is symmetrical about the midpoint of the time interval $[0,T]$, the effective mass is given by (59) and the shift function is given by (60).

The state in expression (90) consists of a superposition of states, where each one of them has started with the particle having a sharp value $x'$ of the position. Subsequently the particle has evolved under the potential $V(x)$, undisturbed by the pointer. At the same time the pointer has evolved as a free wavepacket with an effective mass, and at the start of the measurement the wavepacket is centered at $X_0 = s(x,x',T)$.

Finally we can easily get the state of the system at any time during the measurement from unitary evolution (70) by inserting the propagator (67) for the harmonic oscillator, and similarly for the free particle.

## 5. The Probability of the Position of the Particle in the State after the Measurement

We next turn to the probability distribution of the position of the particle in the final state described by (75). We consider the free particle case and rewrite (73)

$$\Phi_{M_{eff}}(X - g\frac{x' + x}{2}, T) = \int_{-\infty}^{\infty} dX' \langle X|U_{M_{eff}}(T,0)|X' + g(x' + x)/2\rangle\langle X'|\Phi_0\rangle , \quad (91)$$

where, with $\hbar = 1$ and $P$ the momentum of the pointer,

$$U_{M_{eff}}(T,0) = \exp\left[-i\left(P^2/2M_{eff}\right)T\right] . \quad (92)$$

This is an effective time evolution operator for the pointer corresponding top a free particle with mass $M_{eff}$. The probability distribution for the position of the particle is obtained by integrating over the pointer coordinate. We obtain

$$\int_{-\infty}^{\infty} dX |\psi(x,X,T)|^2 = \int_{-\infty}^{\infty} dX \left| \int_{-\infty}^{\infty} dx' \langle x|U_{particle}(T,0)|x'\rangle\langle x'|\varphi_0\rangle \cdot \right.$$

$$\left. \int_{-\infty}^{\infty} dX' \langle X|U_{M_{eff}}(T,0)|X' + g(x' + x)/2\rangle\langle X'|\Phi_0\rangle \right|^2 , \quad (93)$$

where

$$\langle x|U_{particle}(0,T)|x'\rangle = \varphi_m(x - x', T) \quad (94)$$

and $\varphi_m(x - x', T)$ is given by (72).

The integration over the pointer variables in (93) yields

$$\int_{-\infty}^{\infty} dX \int \int_{-\infty}^{\infty} dX' \int_{-\infty}^{\infty} dX'' \langle X'' + g(x'' + x)/2|U_{M_{eff}}^{+}(T,0)|X\rangle \cdot$$

$$\langle X|U_{M_{eff}}(T,0)|X' + g(x' + x)/2\rangle\langle \Phi_0|X''\rangle\langle X'|\Phi_0\rangle = \quad (95)$$

$$\int_{-\infty}^{\infty} dX' \Phi_0^*\left(X' + g(x'' - x')/2\right)\Phi_0(X') .$$

This result exhibits the entanglement between particle and pointer with $\Phi_0(X)$ the initial state of the pointer.

Finally we obtain for the probability distribution

$$\int_{-\infty}^{\infty} dX |\psi(x,X,T)|^2 = \int_{-\infty}^{\infty} dx'' \varphi_m^*(x-x'',T)\varphi_0^*(x'') \int_{-\infty}^{\infty} dx' \varphi_m(x-x',T)\varphi_0(x') \cdot \tag{96}$$

$$\int_{-\infty}^{\infty} \Phi_0^*(X + g\frac{x''-x'}{2})\Phi_0(X)dX \ .$$

If we next integrate over the position of the particle, provided the initial state of the system is normalized, we get $\int_{-\infty}^{\infty} dx \int_{-\infty}^{\infty} dX |\psi(x,X,T)|^2 = 1$ as expected by unitarity.

The third factor in (96) above is an overlap integral, which we can expand in powers of $(x'' - x')$, and write

$$\int_{-\infty}^{\infty} \Phi_0^*\left(X + g(x''-x')/2\right)\Phi_0(X)dX \approx \int_{-\infty}^{\infty} \Phi_0^*(X)\Phi_0(X)dX + O(x''-x') \ . \tag{97}$$

If we assume that the initial state of the pointer is a very narrow normalizable wavepacket, then we get a significant contribution to (96) only in the region $x'' \cong x'$ [16], and we can approximate the overlap integral by the first term in (97).
Under these conditions the probability distribution for the position of the particle in the final state is approximately

$$\int_{-\infty}^{\infty} dx \int_{-\infty}^{\infty} dX |\psi(x,X,T)|^2 \approx |\varphi(x,T)|^2 \tag{98}$$

with

$$\varphi(x,T) = \int_{-\infty}^{\infty} dx' \varphi_m(x-x',T)\varphi_o(x') \ . \tag{99}$$

Thus, if the initial state of the pointer is narrow, the entanglement between the particle and the pointer is negligible. Furthermore, the probability distribution for the position of the particle in the state right after the measurement is approximately the same as the probability distribution in the state at time $T$ corresponding to the particle evolving uncoupled from the pointer.

The opposite situation occurs when the initial state of the pointer is very broad, with width $\Delta(0)$. For then

$$\Phi_0\left(X + g(x'' - x')/2\right) \approx \Phi_0(X) + O\left(1/\Delta(0)\right) \tag{100}$$

and the same result (98) follows. However, as mentioned earlier, the pointer now gives no information about the average position of the particle.

Finally in the limit of an instantaneous measurement, expression (90) for the probability distribution of the particle yields the von Neumann result:

$$\int_{-\infty}^{\infty} dX |\psi(x,X,T)|^2 \underset{T \to 0}{=} |\varphi_0(x)|^2 . \tag{101}$$

Also note that expression (96) for the probability distribution is independent of the mass of the pointer.

We can obtain the probability distribution for the harmonic oscillator by simply replacing the free particle propagator $\varphi_m(x - x', T)$ by the propagator (55) for the harmonic oscillator.

Next we consider the non-normalizable initial state where the particle is a sharp position state at $x_0$, and the pointer is a normalized wavepacket $\Phi_0(X)$ centered at $X = 0$:

$$\psi_0(x, X) = \delta(x - x_0)\Phi_0(X) . \tag{102}$$

It follows from (96) that the probability distribution is

$$\int_{-\infty}^{\infty} dX |\psi(x,X,T)|^2 = |\langle x|U_{particle}(T,0)|x_0\rangle|^2 \tag{103}$$

or, substituting for the free particle propagator (36), we obtain a constant probability distribution

$$\int_{-\infty}^{\infty} dX |\psi(x,X,T)|^2 = \frac{m}{2\pi T} \qquad (104)$$

and also for the harmonic oscillator

$$\int_{-\infty}^{\infty} dX |\psi(x,X,T)|^2 = \frac{m\omega}{2\pi \sin \omega T} \ . \qquad (105)$$

As a result of unitarity and of the structure of the particle propagators these distributions are constant at the end of the measurement, when we have started with a non-normalizable initial state. Also, the results are independent of the initial state of the pointer as long as the pointer starts in a normalized state.

We can solve for the Heisenberg equation of motion for the free particle:

$$\hat{x}(t) = \hat{x}(0) + \frac{\hat{p}(0)}{m} T \qquad (106)$$

In the position basis, and with $\hbar = 1$, the eigenstates of $\hat{x}(T)$ are solutions of the equation

$$x \langle x | x(T) \rangle + \frac{T}{m} \frac{1}{i} \frac{\partial}{\partial x} \langle x | x(T) \rangle = x(T) \langle x | x(T) \rangle \ , \qquad (107)$$

solving this equation gives

$$\langle x | x(T) \rangle = A e^{i \frac{m}{T} \left[ xx(T) - x^2/2 \right]} \ . \qquad (108)$$

Imposing the delta function normalization

$$\langle x(T) | x'(T) \rangle = \delta \left( x'(T) - x(T) \right) \qquad (109)$$

yields

$$A = \left(\frac{m}{2\pi T}\right)^{1/2}. \tag{110}$$

Next the normalized eigenstate (108) yields the transition probability for the particle to go from $x(0)$ to $x(T)$

$$\left|\langle x(T)|x(0)\rangle\right|^2 = \frac{m}{2\pi T}, \tag{111}$$

and comparing with result (104), we obtain

$$\int_{-\infty}^{\infty} dX \left|\psi(x, X, T)\right|^2 = \left|\langle x(T)|x(0)\rangle\right|^2 \tag{112}$$

for the probability distribution of the position of the particle in the state after the measurement. The state after the measurement is given by (71), with the pointer giving a distribution for the average value of the position $\bar{x} = (x_0 + x)/2$, where $x_0$ is fixed. Thus the probability of the position of the pointer after the measurement is equal to the transition probability for the particle to go from $x(0) = x_0$ to $x(T)$.

For the harmonic oscillator we get the Heisenberg equation of motion

$$\hat{x}(T) = \frac{\hat{p}(0)}{m\omega}\sin\omega T + \hat{x}(0)\cos\omega T. \tag{113}$$

Proceeding as in the free particle case we obtain the normalized eigenstate of $\hat{x}(T)$:

$$\langle x|x(T)\rangle = \left(\frac{m\omega}{2\pi\sin\omega T}\right)\exp\left\{i\frac{m\omega}{\sin\omega T}\left[xx(T) - \frac{x^2}{2}\cos\omega T\right]\right\}, \tag{114}$$

and the eigenstate (114) yields the transition probability, which is the same as (105)

$$\left|\langle x(T)|x(0)\rangle\right|^2 = \frac{m\omega}{2\pi\sin\omega T}, \tag{115}$$

thus obtaining the same result (112).

The state after the measurement is

$$\psi(x,X,T) = K_{SHO}(x,x_0,T)\Phi_{M_{eff}}\left(X - g\frac{\tan(\omega T/2)}{\omega T/2}\frac{x_0 + x}{2}\right) \ , \tag{116}$$

where $K_{SHO}(x,x_0,T)$ is the propagator for the harmonic oscillator. The probability of the pointer indicating the average position is the same as the transition probability for the particle to go from $x(0) = x_0$ to $x(T)$.

If, on the other hand, the initial state of the pointer is a sharp position state at $X = 0$ and the particle is in a normalized wavepacket $\varphi_0(x)$, then we obtain for the probability distribution

$$\int_{-\infty}^{\infty}|\psi_0(x,X,T)|^2\,dX = \frac{2}{g}\int_{-\infty}^{\infty}|\langle x|U_{particle}(T,0)|x'\rangle|^2\,|\varphi_0(x')|^2\,dx' . \tag{117}$$

This expression leads to constant distributions both for the free particle and the harmonic oscillator: $m/2\pi T$ and $m\omega/2\pi\sin\omega T$ respectively.

   To conclude this section, we note that in the limit where the mass of the particle is infinite, the position of the particle is a constant of the motion, and a perfect measurement is possible [9], [10]. In this case expression (96) for the probability distribution reduces to the von Neumann result (101) noted earlier, when the duration of the measurement was very short and the pointer indicated the position of the particle in the initial state. The only limitation now comes in the amount of time available for the measurement due to the fact that the pointer is decaying in time, spreading while centered at the position of the particle.

## 6. Summary and Conclusion

We have investigated measuring the position of a moving particle using the von Neumann measurement process extended over a finite period of time. Under these circumstances the motion of the particle and the pointer cannot be neglected. We have examined two cases in detail: ( i ) a free particle and a pointer of finite mass when the coupling function in the interaction (1) is a constant during the measurement; ( ii ) a simple harmonic oscillator when the pointer has infinite mass and the coupling function is symmetrical about the midpoint of the time interval for the measurement. We have found that, if we start with a particle state with a definite value of the position and the pointer state represented by a wavepacket centered at the origin, then, at the end of the measurement, the particle has evolved undisturbed by the coupling with the pointer. The pointer, on the other hand, has spread like a free particle with an effective mass $M_{eff}$, which in general depends on the masses of the particle and the pointer, the duration of the measurement, and the form of the potential acting on the particle. Based on the general structure of the time evolution operator for this system, a qualitative argument has been given to suggest that this free particle behavior of the pointer may be expected in general for any smooth potential $V(x)$ acting on the particle. For the free particle case (i) we have $M_{eff} = M/(1 + g^2 M/12m)$, which combines the mass $M$ of the pointer and the mass $m$ of the particle. The pointer has shifted its center to indicate the average between the positions of the particle in the initial state and in the final state. In this case, for a pointer represented by a Gaussian wavepacket, the spread is given by $\Delta(T) = L(1 + 4T^2/M_{eff}^2 L^4)^{1/2}$, where $L$ is the initial spread. This represents a rate of spreading that is more than in the case of no coupling. For case (ii) there would be no spreading if the coupling were zero. Instead, with the coupling turned on, with the coupling function being a constant, the pointer acquires the effective mass $M_{eff} = m\omega^2 T^2 \Big/ g^2 \left[ \dfrac{\tan(\omega T/2)}{\omega T/2} - 1 \right]$, which produces spreading.

After the measurement the pointer has shifted its center to indicate the average between the positions of the particle in the initial and final state. In both cases the shift is independent of the mass of the particle. For case ( i ) the shift, or indicator function, is given by $s(x, x_0, T) = g(x + x_0)/2$, where $x_0$ is the initial position of the particle and $x$ is the position in the final state. For case ( ii ), with a constant coupling function, we get the shift $s(x, x_0, T) = g \frac{\tan(\omega T/2)}{\omega T/2} \left( \frac{x + x_0}{2} \right)$, which puts a limit on the duration of the measurement: $T < T_{SHO}/2$, where $T_{SHO}$ is the period of the oscillator.

The state of the particle keeps changing during the measurement, and the probability distribution for the position in the initial state cannot be obtained from the state of the system at the end of the measurement. That is, the position of the particle does not commute with the Hamiltonian of the system, and this precludes a perfect measurement. During the measurement, in case ( i ) the particle evolves freely; and the pointer, instead of shifting instantaneously, shifts gradually at constant velocity $V = g\bar{x}/T$, until the measurement is completed, whereupon it indicates the full value of the average position. In case ( ii ) the pointer accelerates with increasing acceleration until it shows the full value above. In addition, in both cases we find the following result: the probability for the pointer to indicate a given average position at the end of the measurement is a constant and equal to the transition probability of the particle to go from the starting eigenstate of the position at $x_0$ to the eigenstate of the position at $x(T)$, where $x(T)$ is an eigenvalue of the Heisenberg position operator.

When the initial state of the particle is not a position eigenstate, then the final state of the system is a superposition of states of the form described above. Each one of these states corresponds to the particle being in a position eigenstate at the start of the measurement. In this case, whenever an observation is made, the pointer can indicate one or another of the different average positions available to the particle. In the limit of an instantaneous measurement we recover the familiar von Neumann result, with the pointer indicating the position of the particle in the initial state.

We can gain some insight into this measurement for the free particle case if the initial state of the particle is a narrow Gaussian wavepacket of spread $l$, and the duration of the measurement is $T$, such that $T < ml^2/2$. In the case of a Gaussian pointer, with an initial spread $L$, must have an effective mass such that $M_{eff} > ml^2/\sqrt{3}L^2$ so that the spread of the pointer has not doubled by the time the measurement is completed. The pointer will indicate an average position within $(x_1 + x_2)/2 \pm \Delta$, where $x_1$ and $x_2$ are the positions of the center of the Gaussian wavepacket describing the particle at the start and at the end of the measurement respectively, and $\Delta \cong l + T^2/m^2l^3$.

Using unitarity, we have obtained and expression (96) for the probability distribution for the position of the particle in the final state of the system, both for the case of the free particle and for the harmonic oscillator. This expression displays the entanglement between the particle and the pointer in the form of an overlap integral over the pointer coordinate involving the initial state of the pointer only. If the initial state of the pointer is given by a narrow wavepacket, then the probability distribution for the pointer in the state immediately after the measurement is approximately the same as the probability distribution obtained by letting the particle evolve uncoupled from the pointer. This probability distribution is evaluated at the end of the measurement rather than at the beginning. In this sense this is an almost "perfect" measurement.

For the case of a particle starting at an eigenstate of the position, this result is exact, regardless of the initial state of the pointer, and it agrees with the result for the transition probability mentioned earlier. Also, in the limit of a very short measurement, the exact expression (96) reduces to the von Neumann result (5).

To conclude, when the measurement has finite duration both the motion of the pointer and the particle influence the result of the measurement. Provided that the particle has a sharp value of its position at the start of the measurement, the pointer now indicates the arithmetic average between the initial and final position of the particle, both for the free particle with finite mass pointer and for the harmonic oscillator with infinite mass pointer when the coupling function in (1) is symmetrical about the center of the time interval for the measurement. Furthermore, the probability that the pointer will indicate a given average value is equal to the transition probability for the undisturbed particle to experience the change in position. If the initial state of the pointer is a narrow wavepacket, then for any initial state of the particle, the measurement yields, approximately, the undisturbed probability distribution for the position of the particle at the end of the measurement.

## Acknowledgement

I wish to thank James B. Hartle for very useful conversations during the course of this work and for his comments and suggestions on the manuscript

# References

[1]  J. von Neumann, *Mathematical Foundations of Quantum Mechanics* (Princeton University Press, Princeton, 1955).

[2]  L.E. Ballentine, *Quantum Mechanics* (Prentice Hall, New Jersey, 1999)

[3]  M. B. Mensky, Phys. Rev. D**20**, 384 (1979).

[4]  C. M. Caves, Phys. Rev. A**36**, 5543 (1987).

[5]  A. Peres and W.K. Wooters, Phys. Rev. D**32**, 1968 (1985).

[6]  D. Sokolovski, Phys. Rev. A**57**, R1469 (1998).

[7]  D. Sokolovski, Phys. Rev. A**59**, 1003 (1999).

[8]  Y. Liu and D. Sokolovski, Phys. Rev. A**63**, 014102/1-4 (2001).

[9]  H. Araki and M. M. Yanase, Phys. Rev. **120**, 622 (1960).

[10] G.C. Ghiradi, F. Minglietta, A. Rimini and T. Weber, Phys. Rev. D**24**, 347 (1981).

[11] J. B. Hartle, in *Quantum Cosmology and Baby Universes: Proceedings of the 1989 Jerusalem Winter School for Theoretical Physics*, edited by S. Coleman, J. B. Hartle, T. Piran, and S. Weinberg (World Scientific, Singapore, 1991), pp. 65-157.

[12] J. B. Hartle, Phys. Rev. D**44**, 3173 (1991).

[13] R. J. Micanek and J. B. Hartle, Phys. Rev. A**54**, 3795 (1996).

[14] R. P. Feynman and A. R. Hibbs, *Quantum Mechanics and Path Integrals* (McGraw-Hill, New York, 1965)

[15[ H. Kleinert, *Path Integrals in Quantum Mechanics, Statistics, and Polymer Physics*, 2$^{nd}$ ed. (World Scientific, New Jersey, 1995)

[16] In fact, for the Gaussian (85) with small spread $L$, one gets for the overlap integral the narrow distribution $\exp\left[-g^2(x'' - x')^2/8L^2\right]$ around $x'' = x'$.